\documentclass[11pt]{article}
\usepackage{latexsym,amsmath,amsfonts,amssymb}
\usepackage{jheppub}

\def\CA{{\cal A}}

\def\CC{{\cal C}}

\def\CL{{\cal L}}

\def\CN{{\cal N}}

\def\CP{{\cal P}}

\def\CS{{\cal S}}

\def\BC{\mathbb{C}}
\def\BH{\mathbb{H}}

\def\BM{\mathbb{M}}
\def\BP{\mathbb{P}}

\def\BS{\mathbb{S}}

\makeatother

\begin{document}
%
\preprint{YITP-14-78, UT-14-40}
\title{Supersymmetric R{\'e}nyi Entropy in Five Dimensions}

\author[a]{Naofumi Hama,}
\author[b]{Tatsuma Nishioka}
\author[c]{and Tomonori Ugajin}

\affiliation[a]{Yukawa Institute for Theoretical Physics, Kyoto University, \\
Kyoto 606-8502, Japan}
\affiliation[b]{Department of Physics, Faculty of Science,
The University of Tokyo, \\
Bunkyo-ku, Tokyo 113-0033, Japan}
\affiliation[c]{Kavli Institute for Theoretical Physics, University of California\\
Santa Barbara, CA 93106-4030, USA}

\emailAdd{hama@yukawa.kyoto-u.ac.jp}
\emailAdd{nishioka@hep-th.phys.s.u-tokyo.ac.jp}
\emailAdd{ugajin@kitp.ucsb.edu}

\abstract{
We introduce supersymmetric R{\'e}nyi entropies for $\CN=1$ supersymmetric gauge theories in five dimensions.
The matrix model representation is obtained using the localization method and the large-$N$ behavior is studied.
The gravity dual is a supersymmetric charged topological AdS$_6$ black hole in the Romans $F(4)$ supergravity.
The variation of the supersymmetric R{\'e}nyi entropy due to the insertion of a BPS Wilson loop is computed.
We find perfect agreements between the large-$N$ and the dual gravity computations both with and without the Wilson loop operator.
}

\maketitle

\section{Introduction}
An intriguing relation between entanglement entropy and partition functions holds in quantum field theories \cite{Calabrese:2004eu}.
The R{\'e}nyi entropy $S_n$, one parameter generalization of  entanglement entropy $S=S_1$, can be written as
\begin{align}\label{REdef}
S_n = \frac{1}{1-n} \log \left[ \frac{Z_n}{(Z_1)^n} \right] \ ,
\end{align}
where $Z_n$ is the partition function on the $n$-covering space $\BM_n$ around an entangling surface $\Sigma$.
If quantum field theories in $d$ dimensions have conformal symmetry and the entangling surface is a $(d-2)$-dimensional hypersphere, $\Sigma=\BS^{d-2}$, $\BM_n$ becomes the $n$-covering space of a $d$-sphere \cite{Casini:2011kv}.
Especially, the entanglement entropy $S$ obtained in the $n\to 1$ limit is equal to the free energy on $\BS^d$, $S = \log Z_1\equiv -F$.
This relation makes it easier to calculate the R{\'e}nyi entropy for free fields \cite{Klebanov:2011uf}, but cannot be applied to supersymmetric theories because the conical singularity at $\Sigma$ breaks supersymmetries.
To remedy the situation, a supersymmetric extension of the R{\'e}nyi entropy
was introduced in \cite{Nishioka:2013haa} by modifying \eqref{REdef} to
\begin{align}\label{SRE_def}
S_n^\text{susy} = \frac{1}{1-n} \log \left| \frac{Z_n^\text{susy}}{(Z_1)^n} \right| \ ,
\end{align}
where $Z_n^\text{susy}$ is the supersymmetric partition function on the $n$-covering $d$-sphere with an $R$-symmetry background field to preserve supersymmetries.
$\CN=2$ superconformal field theories in $d=3$ dimensions are considered in \cite{Nishioka:2013haa} and the matrix model representation of the supersymmetric R{\'e}nyi entropies are derived by using the localization method \cite{Kapustin:2009kz,Jafferis:2010un,Hama:2010av,Hama:2011ea,Imamura:2011wg}.
The gravity dual of the supersymmetric R{\'e}nyi entropy is constructed in \cite{Huang:2014gca,Nishioka:2014mwa} and turns out to be a supersymmetric AdS$_4$ charged topological black hole. The holographic computation precisely agrees with the large-$N$ limit of the supersymmetric R{\'e}nyi entropy both with and without Wilson loop operators.
Similar constructions have been carried out recently in $d=4$ dimensions by \cite{Huang:2014pda,Crossley:2014oea} extending the works \cite{Pestun:2007rz,Hama:2012bg} for $\CN=2$ supersymmetric gauge theories on $\BS^4$.
The four-dimensional supersymmetric R{\'e}nyi entropy has a logarithmic divergence with a coefficient determined by the $a$-anomaly of the Weyl symmetry.

The objective of this paper is to introduce the supersymmetric R{\'e}nyi entropy \eqref{SRE_def} for $\CN=1$ supersymmetric gauge theories in five dimensions and derive their matrix model representations by the localization technique.
We will construct the theories by taking the rigid limit \cite{Festuccia:2011ws} of the $\CN=1$ five-dimensional supergravity \cite{Kugo:2000hn,Kugo:2000af}.
In some aspects, this can be regarded as a theoretical challenge to define field theories on a singular manifold with supersymmetry preserved.
The partition functions of supersymmetric gauge theories are calculated on a round five-sphere in \cite{Kallen:2012cs,Hosomichi:2012ek,Kallen:2012va,Kim:2012ava,Kallen:2012zn,Jafferis:2012iv} and on a squashed five-sphere in \cite{Imamura:2012xg,Imamura:2012bm,Lockhart:2012vp,Kim:2012qf}.
More general five-dimensional manifolds admitting rigid supersymmetry are explored in \cite{Kawano:2012up,Kim:2012gu,Terashima:2012ra,Fukuda:2012jr,Yagi:2013fda,Lee:2013ida,Cordova:2013cea,Pan:2013uoa,Imamura:2014ima}.
We will show that the Killing spinor equations and additional equations for the Killing spinors can be solved on the resolved space of the $n$-covering five-sphere.

Using the supersymmetry generated by the solution of the Killing spinor equations, we perform the localization computation for the partition function on the resolved sphere that is the Hopf fibration over deformed $\BC\BP^2$.
There are three fixed points on $\BC\BP^2$ for $U(1)^2$ actions inside  $U(1)\times SO(4)$ symmetry the resolved sphere has.
We notice that the resolved five-sphere can be identified with the squashed five-sphere with the squashing parameters $(\omega_1, \omega_2, \omega_3)= (1/n, 1, 1)$ near the fixed points.
Translating the results for the squashed sphere \cite{Imamura:2012xg,Imamura:2012bm,Lockhart:2012vp,Kim:2012qf}, we obtain the perturbative partition function on the $n$-covering five-sphere.

We evaluate the supersymmetric R{\'e}nyi entropy \eqref{SRE_def} in the large-$N$ limit of $\CN=1$ $USp(2N)$ superconformal gauge theories and reveal the $n$-dependence that satisfies several inequalities the usual R{\'e}nyi entropy does.
We also consider an addition of a Wilson loop preserving supersymmetry, that can be physically interpreted as an insertion of a quark inside an entangling surface \cite{Lewkowycz:2013laa}.
The variation of the supersymmetric R{\'e}nyi entropy by the loop turns out to be independent of the parameter $n$.

Finally, we construct the gravity dual of five-dimensional supersymmetric R{\'e}nyi entropy as a solution of the Romans $F(4)$ supergravity \cite{Romans:1985tw}.
It is a  supersymmetric charged topological black hole in the Euclidean AdS$_6$ space whose boundary is $\BS^1 \times \BH^4$.\footnote{Non-supersymmetric charged topological black holes are studied as gravity duals of charged R{\'e}nyi entropies in \cite{Belin:2013uta,Belin:2013dva,Pastras:2014oka,Belin:2014mva}.}
We compute the holographic free energy following \cite{Alday:2014rxa,Alday:2014bta} and the expectation value of the holographic Wilson loop.
We find that the holographic supersymmetric R{\'e}nyi entropy and its variation due to the loop perfectly agree with the results of the dual field theory in the large-$N$ limit.

\bigskip
\noindent
{\bf Note added:} While this work was being completed, we were aware of the paper \cite{Alday:2014fsa}, which has a substantial overlap with this paper.

\section{Rigid $\CN=1$ supersymmetry in five dimensions}

The $\CN=1$ supergravity coupled to Yang-Mills and matter fields in five dimensions is constructed in \cite{Kugo:2000hn}.
This theory has an $SU(2)_R$ $R$-symmetry and the Weyl multiplet consists of the vielbein $e^a_\mu$, the graviphoton $\CA_\mu$, the $SU(2)_R$ gauge field $V_\mu^{IJ}$, the $SU(2)_R$ triplet scalar field $t^{IJ}$, the dilaton $\alpha$, the real anti-symmetric tensor $v_{ab}$, the real scalar $C$, 
the $SU(2)_R$-Majorana gravitino $\psi^I_\mu$ and $SU(2)_R$-Majorana fermion $\chi^I$.

Since the dilaton does not change under the supersymmetry transformation, we can fix the dilatational symmetry by $\alpha=1$.
In addition, we only consider theories without central charge which allows us to {\it turn off the graviphoton $\CA_\mu=0$}.
This simplifies the construction of supersymmetric field theories on a curved space from the $\CN=1$ supergravity as we will perform below.

Following \cite{Festuccia:2011ws}, we take the rigid limit of the $\CN=1$ supergravity by setting the gravitino $\psi_\mu^I$ and the fermion $\chi^I$ and their variations to be zero while letting the space-time be curved:
\begin{align}\label{KSE}
\begin{aligned}
\delta \psi_\mu^I &= \nabla_\mu \xi^I - i t^I_{~J} \Gamma_\mu \xi^J - (V_\mu)^I_{~J} \xi^J + \frac{i}{2}v^{\nu\rho}\Gamma_{\mu\nu\rho} \xi^I 
= 0 \ , \\
\delta \chi^I &= \frac{i}{2}\Gamma^\mu \xi^I \nabla^\nu v_{\mu\nu} + \frac{i}{2}D_\mu t^I_{~J} \Gamma^\mu \xi^J + 
v_{\mu\nu} 
\Gamma^{\mu\nu} t^I_{~J}\xi^J + \frac{1}{2}C\xi^I 
= 0\  ,
\end{aligned}
\end{align}
where $\nabla_\mu$ is the covariant derivative with respect to the Lorentz index, 
and 
\begin{align}
\begin{aligned}
\nabla_\mu v_{ab} &= \partial_\mu v_{ab} + \omega_{\mu\, a}^{~~~c} v_{cb} - \omega_{\mu\, b}^{~~~c} v_{ca} \ , \\
D_\mu t^{I}_{~J} &= \partial_\mu t^{I}_{~J} - (V_{\mu})_{~K}^{I} t^{K}_{~J} + (V_{\mu})^{K}_{~J} t^{I}_{~K} \ .
\end{aligned}
\end{align}
In this limit, the supersymmetry transformations of the other background fields automatically vanish.

\subsection{Supersymmetry algebra and multiplets}

The $\CN=1$ supersymmetry algebra in five dimensions is given by
\begin{align}\label{SUSYAlgebra}
\begin{aligned}
	\{ \delta_{\xi_1} , \delta_{\xi_2} \} = v^\mu D_\mu + \delta_M(\Theta_{ab}) + \delta_R(R^{IJ})  + \delta_G(\gamma)\ ,
\end{aligned}
\end{align}
where $D_\mu$ is the covariant derivative with respect to the gauge symmetry and translation, $v_\mu$, $\Theta_{ab}$, $R^{IJ}$ and $\gamma$ are parameters for the translation, Lorentz rotation, $SU(2)_R$-symmetry and gauge symmetry transformations\footnote{We contract the $SU(2)_R$ indices from northwest to southeast direction when the indices are suppressed.}
\begin{align}
\begin{aligned}
	v_\mu &= 2\xi_1 \Gamma_\mu \xi_2 \ , \\
	\Theta_{ab} &= 2i (\xi_1 \Gamma_{abcd}\xi_2) v^{cd} - 2i(\xi_1^I\Gamma_{ab}\xi_2^J + \xi_1^J\Gamma_{ab}\xi_2^I)t_{IJ} \ , \\
	R^{IJ} &= 6i (\xi_1 \xi_2) t^{IJ} +2i (\xi_1^I\Gamma_{ab}\xi_2^J + \xi_1^J\Gamma_{ab}\xi_2^I)v^{ab} \ , \\
	\gamma &= -2i (\xi_1 \xi_2) \sigma \ .
\end{aligned}
\end{align}

\paragraph{Vector multiplet.}
The vector multiplet contains the gauge field $A_\mu$, a real scalar $\sigma$, an $SU(2)_R$ triplet scalar $Y_{IJ}$ and an $SU(2)_R$-Majorana fermion $\lambda^I$.
They transform under the supersymmetry as\footnote{We can redefine the triplet scalar $Y_{IJ}$ in \cite{Zucker:1999ej,Kugo:2000hn,Kugo:2000af,Fujita:2001kv} as $D_{IJ} = 2Y_{IJ} - 2t_{IJ} \sigma$ to make contact with \cite{Hosomichi:2012ek}.}
\begin{align}\label{VM_transform}
\begin{aligned}
\delta A_\mu &= -2 \xi\Gamma_\mu\lambda \ , \\
\delta \sigma &= 2i \xi \lambda \ , \\
\delta \lambda^I &= \frac{1}{4}\Gamma^{\mu\nu} \xi^I 
F_{\mu\nu}
- \frac{i}{2}\Gamma^\mu \xi^I D_\mu \sigma - Y^{IJ}\xi_J  \ , \\
\delta Y^{IJ} &= - \xi^{I}\Gamma^\mu D_\mu \lambda^{J} + \frac{i}{2} v^{\mu\nu} \xi^{I}\Gamma_{\mu\nu}\lambda^{J} + i t^J_{~K}  \xi^I \lambda^K + 2i\, t^{IJ}  \xi \lambda 
-i   [\sigma, \xi^I \lambda^J] + (I\leftrightarrow J) \ .
\end{aligned}
\end{align}
Here $F_{\mu\nu} = \nabla_\mu A_\nu - \nabla_\nu A_\mu - [A_\mu, A_\nu]$.
The covariant derivatives are
\begin{align}
\begin{aligned}
D_\mu \sigma &= \partial_\mu \sigma - [A_\mu, \sigma] \ , \\
D_\mu \lambda^I &= \nabla_\mu \lambda^I - (V_\mu)^I_{~J} \lambda^J - [A_\mu , \lambda^I] \ .
\end{aligned}
\end{align}

To put them into cohomological forms, we redefine the gaugino as \cite{Kallen:2012cs}
\begin{align}
\lambda^I = \frac{1}{4}\left( -\xi^I v^\mu\Psi_\mu - \Gamma^\mu\xi^I \Psi_\mu - \Gamma^{\mu\nu}\xi^I \chi_{\mu\nu} \right) \ ,
\end{align}
where 
\begin{align}
\Psi_\mu = -2\xi\Gamma_\mu \lambda \ , \qquad \chi_{\mu\nu} =  \xi\Gamma_{\mu\nu} \lambda\ .
\end{align}
Then the supersymmetry transformation law \eqref{VM_transform} becomes
\begin{align}
\begin{aligned}
\delta A_\mu &= \Psi_\mu \ , \\
\delta \Psi_\mu &= v^\nu F_{\nu\mu} + i D_\mu \sigma \ , \\
\delta \sigma &= -i v^\mu \Psi_\mu \ , \\
\delta \chi_{\mu\nu} &\equiv H_{\mu\nu}=\frac{1}{4}(\xi^I \Gamma_{\mu\nu\rho\sigma}\xi_I)F^{\rho \sigma}-\frac{1}{2}F_{\mu\nu}-\frac{i}{2}(v_\mu D_\nu \sigma - v_\nu D_\mu \sigma)+ (\xi^I \Gamma_{\mu\nu}\xi^J)Y_{IJ} \ , \\
\delta H_{\mu\nu} &=  v^\rho D_\rho \chi_{\mu\nu} + i [\sigma, \chi_{\mu\nu}] +\Theta_\mu {}^\rho \chi_{\nu\rho}-\Theta_\nu {}^\rho \chi_{\mu\rho}\ . 
\end{aligned}
\end{align}

The supersymmetric action of the vector multiplet is
\begin{align}
\begin{aligned}
\CL_\text{YM} &= - \frac{2}{g^2}\text{Tr} \bigg[ \frac{1}{4} F_{\mu\nu}F^{\mu\nu} - \frac{1}{2}D_\mu \sigma D^\mu \sigma - Y_{IJ}Y^{IJ} + 2 \sigma (2 t_{IJ}Y^{IJ} -  F_{\mu\nu}v^{\mu\nu}) +2(C-4t_{IJ}t^{IJ})\sigma^2 \\
	&\qquad \qquad \quad +2\lambda \Gamma^\mu D_\mu \lambda - i \lambda^I (\epsilon_{IJ} \Gamma_{\mu\nu}v^{\mu\nu} - 2t_{IJ})\lambda^J - 2i \sigma [\lambda, \lambda] \bigg] \ ,
\end{aligned}
\end{align}
where $g$ is the gauge coupling constant.
The Chern-Simons term can be added in five dimensions\footnote{Note that there is no imaginary unit as an overall factor because the gauge field $A$ is anti-hermitian in our convention.}
\begin{align}
\CL_\text{CS} (A) &= \frac{k}{24\pi^2} \text{Tr} \left[A\wedge dA \wedge dA + \frac{3}{2}A \wedge A\wedge A \wedge dA + \frac{3}{5} A\wedge A\wedge A\wedge A\wedge A\right] \ ,
\end{align}
whose supersymmetric completion is \cite{Kallen:2012cs}
\begin{align}
\CL_\text{SCS} = \CL_\text{CS}(A - i\sigma \kappa) - \frac{k}{8\pi^2} \text{Tr} \left[ \Psi \wedge \Psi \wedge \kappa \wedge F(A - i\sigma \kappa) \right] \ ,
\end{align}
where $\kappa$ and $\Psi$ are one-forms dual to the Killing vector $\kappa\equiv v_\mu dx^\mu$ and $\Psi\equiv \Psi_\mu dx^\mu$, respectively.

\paragraph{Hypermultiplet.}
We can realize the supersymmetry algebra only for on-shell hypermultiplets. 
We, however, need one supercharge $\delta$ with unit norm $\xi\xi = 1$ for localization which satisfies
\begin{align}\label{PSUSYAlg}
	\delta^2 = v^\mu D_\mu + \delta_M(\Theta_{ab}) + \delta_R(R^{IJ})  + \delta_G(\gamma) \ ,
\end{align}
where 
\begin{align}
\begin{aligned}
	v_\mu &= \xi \Gamma_\mu \xi \ , \\
	\Theta_{ab} &= i (\xi \Gamma_{abcd}\xi) v^{cd} - 2i(\xi^I\Gamma_{ab}\xi^J) t_{IJ}   \ , \\
	R^{IJ} &= 3i t^{IJ} +2i (\xi^I\Gamma_{ab}\xi^J)v^{ab} \ , \\
	\gamma &= -i  \sigma \ ,
\end{aligned}
\end{align}
with the modified supersymmetry transformation law by introducing an auxiliary field \cite{Hosomichi:2012ek}.
They satisfy
\begin{align}
v^\mu\Gamma_\mu \xi_I = \xi_I \ , \qquad v_\mu v^\mu = 1 \ , \qquad  v^a \Theta_{ab} = 0 \ , \qquad \nabla_\mu v_\nu = -\Theta_{\mu\nu} \ .
\end{align}

The hypermultiplet consists of a scalar $q_A^I$, a fermion $\psi_A$ and an auxiliary field $F^A_I$ with flavor indices $A=1,2,\cdots, 2r$ for $r$ hypermultiplets.
The index $A$ is raised and lowered with a $2r\times 2r$ antisymmetric matrix $\Omega_{AB}$ as
\begin{align}
q^I_A = q^{IB} \Omega_{BA} \ , \qquad q^{IA} =  \Omega^{AB} q^I_B \ , \qquad\Omega^{AB}\Omega_{BC} = - \delta^A_C \ .
\end{align}
The reality conditions are imposed by
\begin{align}\label{reality_condition}
\begin{aligned}
(q^I_A)^\ast &= \epsilon_{IJ} \Omega^{AB} q^J_B \ , \qquad
 (\psi_{A}^\alpha)^\ast = \Omega^{AB} \psi_B^\beta \CC_{\beta\alpha} \ , \\
(F^I_A)^\ast &= \epsilon_{IJ} \Omega^{AB} F^J_B \ , \qquad \Omega_{AB} = \Omega^{AB} \ .
\end{aligned}
\end{align}
The generators of the Lie group have one upper and one lower indices of the flavor symmetry, $\sigma^A_{~B}$ for example.

Choosing the invariant tensor $\Omega$ of $Sp(2r)$ to be 
\begin{align}
\Omega_{AB} = \Omega^{AB} = 
\left( \begin{array}{cc}
0 & {\bold 1}_{r} \\
-{\bold 1}_r & 0
\end{array}\right) \ ,
\end{align}
the scalar field $q^I_A$ satisfying the reality condition \eqref{reality_condition} is written as a two-component vector 
\begin{align}
q^1 = \left(\begin{array}{c}
\phi_+ \\
\phi_-
\end{array}\right) \ , \qquad
q^2 = \left(\begin{array}{c}
-\phi_-^\ast \\
\phi_+^\ast
\end{array}\right) \ ,
\end{align}
where $\phi_+$ and $\phi_-$ are in the fundamental and anti-fundamental representation of $SU(r)$ respectively.
The fermion $\psi_A$ is similarly decomposed to 
\begin{align}
\psi_A^\alpha = \left( \begin{array}{c}
\psi_{\hat A}^\alpha \\
- \psi^\ast_{\alpha \hat B}
\end{array}\right) \ ,
\end{align}
as a vector with spinors $\psi_{\hat A}$ in the fundamental representation of $SU(r)$.

The off-shell supersymmetry transformations realizing \eqref{PSUSYAlg} are found to be
\begin{align}\label{HypSusyTr}
\begin{aligned}
\delta q^I_A &= 2i  \xi^I \psi_A \ , \\
\delta \psi_A &= - i(D_\mu q^I_A)  \Gamma^\mu \xi_I - 3t^I_{~J}\xi_I q^J_A 
+ \xi^I \sigma_{AB} q_I^B - v_{\mu\nu} \Gamma^{\mu\nu} \xi_I q^I_A + F^I_A {\check \xi}_I\ , \\
\delta F^I_A &= -2 {\check \xi}^I \left( \Gamma^\mu D_\mu \psi_A + \frac{i}{2} v_{\mu\nu} 
\Gamma^{\mu\nu} \psi_A + i \sigma_{AB} \psi^B + 2i (\lambda^J)_{AB} q_J^B \right) \ ,
\end{aligned}
\end{align}
where the checked parameter $\check \xi^I$ satisfies \cite{Hosomichi:2012ek}
\begin{align}
\begin{aligned}
	\xi^I \xi_I = {\check \xi}^I {\check \xi}_I \ , \qquad \xi_I {\check \xi}_J = 0 \ , \qquad \xi^I\Gamma_\mu \xi_I + {\check \xi}^I\Gamma_\mu {\check \xi}_I = 0 \ ,
\end{aligned}
\end{align}
and the covariant derivatives are
\begin{align}
\begin{aligned}
D_\mu q^I &= \partial_\mu q^I - (V_\mu)^I_{~J} q^J -  A_\mu q^I \ , \\
D_\mu \psi &= \nabla_\mu \psi -  A_\mu \psi \ .
\end{aligned}
\end{align}

We rewrite the transformation laws \eqref{HypSusyTr} with the fermionic variables \cite{Kallen:2012va}
\begin{align}
q_A \equiv \xi_I q^I_A \ , \qquad \psi_A^\pm \equiv P_\pm \psi_A \ ,
\end{align}
which satisfy the ``chirality" conditions $P_+ q_A = q_A$ with the projection operators $P_\pm \equiv \frac{1\pm v^\mu \Gamma_\mu}{2}$.
For a shifted auxiliary field $\tilde F_A \equiv \check \xi_I \tilde F^I_A$ ($P_- \tilde F_A = \tilde F_A$), 
the supersymmetry transformations of the fields $(q_A, \psi_A^\pm, \tilde F_A)$ are recast in the following form:
\begin{align}
\begin{aligned}
	\delta q_A &= -i \psi_A^+ \ , \\
	\delta \psi_A^+ &= i \left( v^\mu D_\mu \delta_A^B + i \sigma_A^{~B} \right) q_B  + \frac{i}{4}\Theta_{ab}\Gamma^{ab} q_A  \ , \\
	\delta \psi_A^- &= \tilde F_A \ , \\
	\delta \tilde F_A &= \left( v^\mu D_\mu \delta_A^B + i  \sigma_A^{~B} \right)\psi_B^- + \frac{1}{4}\Theta_{ab}\Gamma^{ab} \psi_A^-\ .
\end{aligned}
\end{align}

The matter lagrangian reads \cite{Kugo:2000af}\footnote{Although our off-shell supersymmetry transformation \eqref{HypSusyTr} differs from theirs in \cite{Kugo:2000af}, the off-shell lagrangian still closes.
}
\begin{align}
\begin{aligned}
\CL_\text{matter} &= D_\mu \bar q D^\mu q + \left(v_{\mu\nu}v^{\mu\nu} + 2t_{IJ}t^{IJ} - C - \frac{R}{4}\right)\bar q q - \bar q^I (t_{IK}t^K_{~J} + 2\sigma t_{IJ} - \sigma^2 \epsilon_{IJ} - 2Y_{IJ})q^J \\ 
&\qquad + 2\bar\psi \Gamma^\mu D_\mu \psi + i \bar\psi (\Gamma_{\mu\nu} v^{\mu\nu} + 2 \sigma )\psi  - 8i \bar q \lambda \psi - \bar F F \ ,
\end{aligned}
\end{align}
where the flavour indices $A,B$ are contracted from northeast to southwest.

\subsection{Resolved space}

We are interested in $\CN=1$ supersymmetric field theories on a branched $n$-covering of five-sphere. To treat the conical singularity we replace it with a resolved space whose metric is given by
\begin{align}\label{Resolved}
ds^2 = \frac{d\theta^2}{f(\theta)} + n^2 \sin^2\theta d\tau^2 + \cos^2\theta \,ds_{\BS^3}^2 \ ,
\end{align}
where  $f(\theta)$ is a smooth function behaving as 
\begin{align}
f(\theta) = \left\{ \begin{array}{cl}
1/n^2 \ , & \qquad \theta = 0 \ , \\
1 \ , & \qquad \epsilon < \theta < \pi/2 \ ,
\end{array}\right.
\end{align}
for $\epsilon\ll 1$. $ds_{\BS^3}^2$ is the metric of a three-sphere
\begin{align}
ds_{\BS^3}^2 = \sum_{i=1}^3 e_L^i e_L^i \ .
\end{align}
The vielbein $e_L^i$ in the left invariant frame and the spin connections satisfy
\begin{align}
d e^i_L = \epsilon^{ijk}e_L^j \wedge e_L^k \ , \qquad \omega^{ij}_L = \epsilon^{ijk} e_L^k \ .
\end{align}
They can be parametrized by an element $g$ of $SU(2)$ group 
\begin{align}\label{SU2}
 i e_L^i \sigma^i = g^{-1} dg \ ,
\end{align}
with the Pauli matrices $\sigma^i$ $(i=1,2,3)$.
We choose the vielbein of the resolved space \eqref{Resolved} as
\begin{align}\label{Vielbein}
e^1 = \frac{d\theta}{\sqrt{f(\theta)}} \ , \qquad e^2 = n\sin\theta\, d\tau \ , \qquad e^{i+2} = \cos\theta \,e^i_L\quad (i=1,2,3) \ ,
\end{align}
and the spin connections are
\begin{align}\label{SpinConnection}
\omega^{12} = - n\cos\theta \sqrt{f(\theta)}\, d\tau \ ,  \qquad \omega^{1\, i+2} = \sin\theta \sqrt{f(\theta)} \, e^i_L \ , \qquad \omega^{i+2\, j+2} = \omega_L^{ij} \ .
\end{align}

\subsection{Relation between resolved space and squashed five-sphere}\label{ss:Relation}
A five-sphere is embedded into $\BC^3$ by complex coordinates $(z_1,z_2,z_3)$ as $|z_1|^2 + |z_2|^2 + |z_3|^3 =1$.
It has $U(1)^3$ symmetry acting on the coordinates as
\begin{align}\label{S5_sym}
(z_1,z_2,z_3) \to (e^{ia_1}z_1,e^{ia_2}z_2,e^{ia_3}z_3) \ .
\end{align}
The five-sphere has the Hopf fiber representation as the $U(1)$ fibration over $\BC\BP^2$.
The translation along the $U(1)$ fiber is described by the overall $U(1)$ phase rotation $a_1=a_2=a_3$ in \eqref{S5_sym}.
There are three fixed points of $U(1)^2$ symmetry on the base $\BC\BP^2$ at
$(z_1,z_2,z_3) = (1,0,0),\,(0,1,0),\,(0,0,1)$.

Let us introduce new coordinates by
\begin{align}
\begin{aligned}
z_1 &= \sin\theta \,e^{i n \tau} \ , \\
z_2 &= \cos\theta \cos\frac{\phi}{2} \,e^{i(\chi + \xi)/2}  \ , \\
z_3 &= \cos\theta \sin\frac{\phi}{2} \,e^{i(\chi - \xi)/2}  \ ,
\end{aligned}
\end{align}
where the ranges of the angles are taken to be $0\le \theta < \pi/2,\, 0\le \tau < 2\pi, \,0\le\phi< \pi,\, 0\le \chi< 4\pi$ and $0\le \xi < 2\pi$.
The metric is locally that of a five-sphere, but globally the $n$-branched cover 
\begin{align}\label{SingS5}
\begin{aligned}
ds^2 &= d\theta^2 + n^2 \sin^2 \theta d\tau^2 + \cos^2\theta ds_{\BS^3}^2 \ , \\
ds_{\BS^3}^2 &= \frac{1}{4}\left[ d\phi^2 + \sin^2\phi \,d\xi^2 + (d\chi + \cos\phi \,d\xi)^2\right] \ .
\end{aligned}
\end{align}
In this parametrization, the translation along the Hopf fiber is given by the shifts of the angles
\begin{align}
\tau \to \tau + \frac{a}{n} \ , \qquad \chi \to \chi + 2a \ ,
\end{align}
that is generated by a vector field
\begin{align}
v^\mu \partial_\mu = \frac{1}{n}\partial_\tau + 2\partial_\chi \ .
\end{align}
The three fixed points are located at
\begin{align}
\begin{aligned}
(1,0,0): & \quad \theta = \frac{\pi}{2} \ , \\
(0,1,0): & \quad \theta = 0\ , \quad \phi = 0 \ , \\
(0,0,1): & \quad \theta = 0\ , \quad \phi = \pi \ .
\end{aligned}
\end{align}

The vielbein for $\BS^3$ are written as
\begin{align}
\begin{aligned}
e^1_L &= \frac{1}{2}(\sin\phi\cos\chi \,d\xi - \sin\chi \,d\phi) \ , \\
e^2_L &= \frac{1}{2}(\sin\chi \sin\phi \,d\xi + \cos\chi \,d\phi) \ , \\
e^3_L &= \frac{1}{2}(d\chi + \cos\phi \,d\xi) \ .
\end{aligned}
\end{align}

Now we consider a deformation of a five-sphere satisfying $|z_1|^2/n^2+ |z_2|^2 + |z_3|^2   = 1$.
This is a special case of the squashed five-sphere defined by 
\begin{align}
\omega_1^2 |z_1|^2+ \omega_2^2 |z_2|^2 + \omega_3^2 |z_3|^2   = 1 \ ,
\end{align}
with three squashing parameters chosen as
\begin{align}\label{triplesines}
(\omega_1, \omega_2, \omega_3) = \left(\frac{1}{n}, 1, 1\right) \ .
\end{align}

One can parametrize the complex coordinates with the real angles by
\begin{align}
\begin{aligned}
z_1 &= n\sin\theta \,e^{i \tau} \ , \\
z_2 &= \cos\theta \cos\frac{\phi}{2} \,e^{i(\chi + \xi)/2}  \ , \\
z_3 &= \cos\theta \sin\frac{\phi}{2} \,e^{i(\chi - \xi)/2}  \ ,
\end{aligned}
\end{align}
that gives
\begin{align}
ds^2 &= \frac{ d\theta^2}{f_n(\theta)} + n^2 \sin^2 \theta d\tau^2 + \cos^2\theta ds_{\BS^3}^2 \ , 
\end{align}
where $f_n(\theta) = 1/(n^2 \cos^2\theta + \sin^2\theta)$.
This space may be regarded as a resolved space with $f(\theta)=f_n(\theta)$ if the partition function does not depend on the choice of $f(\theta)$.
In appendix \ref{app:OmegaBG}, we show in detail that this identification is possible for evaluating the one-loop partition functions.

\subsection{Killing spinor equations}
We will solve the Killing spinor equations \eqref{KSE} on the resolved space \eqref{Resolved}.
We let spinors $\xi^I$ be tensor products of spinors in two dimensions $\zeta^I$ and spinors in three dimensions $\eta^I$ 
\begin{align}
\xi^I = \zeta^I \otimes \eta^I \ .
\end{align}
Correspondingly, the gamma matrices that are hermitian $(\Gamma^a)^\dagger = \Gamma^a$ can be written in tensor product forms:
\begin{align}
\begin{aligned}\label{5dGamma}
\Gamma^1 &= \sigma^1\otimes {\bf 1}_2 \ ,\qquad \Gamma^2 = \sigma^2 \otimes {\bf 1}_2 \ , \qquad \Gamma^{i+2} = \sigma^3 \otimes \sigma^i\ , \quad (i=1,2,3) \ .
\end{aligned}
\end{align}
The charge conjugation matrix takes the form
\begin{align}
\CC = \sigma_1 \otimes (i\sigma_2) \ .
\end{align}
With the vielbein \eqref{Vielbein} and the spin connections \eqref{SpinConnection}, we find the background fields
\begin{align}
\begin{aligned}
t^1_{~1} =& -t^2_{~2}= \frac{1}{2}\sqrt{f(\theta)}  \ , \qquad v^{12} = \mp i \frac{ \sqrt{f(\theta)} -1 }{2\cos\theta}  \ , \\
(V_\mu)^1_{~1} &= - (V_\mu)^2_{~2} = 
\mp i\frac{n \sqrt{f(\theta)}}{2}  \delta_{\mu \tau} \ ,
\end{aligned}
\end{align}
solve the first line of the Killing spinor equations \eqref{KSE} with the solutions
\begin{align}\label{KSSols}
\begin{aligned}
\xi^1 & = (e^{\frac{i}{2}\theta \sigma_1} 
\zeta^1) \otimes \eta_+ \ , &\qquad &\sigma_3 \zeta^1 = \pm \zeta^1 \ , \\
\xi^2 & = (e^{-\frac{i}{2}\theta \sigma_1} 
\zeta^2) \otimes \eta_- \ , &\qquad &\sigma_3 \zeta^2 = \mp \zeta^2 \ ,
\end{aligned}
\end{align}
where $\zeta^{1,2}$ are constant spinors in two dimensions
and $\eta_\pm$ are the Killing spinors on a unit three-sphere
\begin{align}
\left(\partial_i + \frac{i}{2}\sigma_i\right)\eta_\pm = \pm \frac{i}{2}\sigma_i \eta_\pm \ , \qquad (i=1,2,3) \ .
\end{align}
The $SU(2)$-Majorana condition leads the relations
\begin{align}
\zeta^2 = \sigma_1 \zeta^1 \ , \qquad \eta_- = i\sigma_2 \eta_+ \ ,
\end{align}
which are compatible with the solutions \eqref{KSSols}.

The second line of the Killing spinor equations \eqref{KSE} is satisfied by the solutions \eqref{KSSols} if we choose the scalar field $C$ to be
\begin{align}
C = \frac{1}{4}\cot\theta f'(\theta) - \frac{f(\theta) - \sqrt{f(\theta)}}{2\cos^2\theta} \ .
\end{align}

\section{Localization}
We will localize the infinite-dimensional path integral of the partition function to a finite-dimensional matrix integral by adding a $\delta$-exact term to the action $I \to I + t \delta V$ where the localizing term $\delta V$ is taken to be positive semi-definite.
Since the path integral does not depend on the $\delta$-exact term, we let $t$ be large so that the fixed points are given by $\delta V=0$.
After determining the fixed point loci for the gauge and matter sectors, we will read off the perturbative partition function on the resolved space by identifying it with the squashed five-sphere at the three fixed points on the base $\BC\BP^2$.

\subsection{Gauge sector}
A localization term for the gauge sector is given similarly to \cite{Hosomichi:2012ek} by
\begin{align}
V_\text{gauge} = -4\text{Tr}\left[ (\delta\lambda)^\dagger \lambda\right] \ ,
\end{align}
whose supersymmetry variation yields
\begin{align}\label{Gauge_Loc}
\begin{aligned}
\delta V_\text{gauge}\big|_\text{boson} = - 
\text{Tr}\Bigg[ &\frac{1}{4}\left( F_{\mu\nu} + \frac{1}{2}\varepsilon_{\mu\nu\rho\sigma\kappa} v^\rho F^{\sigma\kappa} \right)\left( F^{\mu\nu} + \frac{1}{2}\varepsilon^{\mu\nu\rho\sigma\kappa} v_\rho F_{\sigma\kappa} \right)  \\
	&\qquad  + \frac{1}{2}(v^\rho F_{\rho\mu})(v_\rho F^{\rho\mu})- D_\mu\sigma D^\mu \sigma + 2 Y_{IJ} Y^{IJ} \Bigg] \ , 
\end{aligned}
\end{align}
where we let the hermitian conjugates of $\sigma$ and $Y_{IJ}$ be $\sigma^\dagger = - \sigma$ and $Y^\dagger_{IJ} = Y^{IJ}$.
The saddle point of the localization term \eqref{Gauge_Loc} is
\begin{align}\label{saddlepts}
F_{\mu\nu} = - \frac{1}{2}\varepsilon_{\mu\nu\rho\sigma\kappa} v^\rho F^{\sigma\kappa} \ , \qquad v^\rho F_{\rho\mu}=0 \ , \qquad D_\mu \sigma = 0 \ , \qquad Y_{IJ} = 0 \ .
\end{align}
In the zero instanton sector, the gauge field is a flat connection and the saddle point becomes
\begin{align}\label{GaugeFP}
A_\mu = 0 \ , \qquad \sigma = \sigma_0 = \text{const}\ ,
\end{align}
up to the gauge transformation.

\subsection{Matter sector}
We choose a localization term for the matter sector as
\begin{align}
V_\text{matter} = (\delta \psi^+_A)^\dagger \psi^{+A} +  (\delta \psi^-_A)^\dagger \psi^{-A} \ ,
\end{align}
whose supersymmetry variation yields
\begin{align}
\begin{aligned}
	\delta V_\text{matter}|_\text{boson} &= \left( \frac{1}{4}\Theta_{ab}\Gamma^{ab} q_A  + v^\mu D_\mu q_A \right)^\dagger \left( \frac{1}{4}\Theta_{ab}\Gamma^{ab} q_A  + v^\mu D_\mu q_A \right) + (\sigma q)^\dagger_A (\sigma q)^A + \tilde F_A^\dagger \tilde F^A \ .
\end{aligned}
\end{align}
Then the path integral localizes to the following fixed locus
\begin{align}
\left( \frac{1}{4}\Theta_{ab}\Gamma^{ab}   + v^\mu D_\mu \right) q_A = 0 \ , \qquad \sigma_{A}^{~B} q_B = 0 \ , \qquad \tilde F_A = 0 \ .
\end{align}
Combining with \eqref{GaugeFP}, one finds
\begin{align}
q_A = \tilde F_A = 0 \ .
\end{align}

\subsection{Partition function on the $n$-covering five-sphere}

As described in section \ref{ss:Relation}, 
the resolved five-sphere \eqref{Resolved} can be regarded as a squashed five-sphere with $f(\theta) = f_n(\theta)$ since they are locally equivalent near the fixed points of the four-dimensional base space in the Hopf fiber representation.
Under this identification, the squashing parameters are $(\omega_1, \omega_2, \omega_3) = (1/n, 1, 1)$ and the perturbative partition function is \cite{Lockhart:2012vp,Imamura:2012bm}
\begin{align}
\begin{aligned}
Z^{{\rm pert}} = \int d \sigma_0 \,e^{-I_0}&  \prod_{\alpha \in {\rm positive \ root}} \mathcal{S}_3\left(\alpha(\sigma_0) \Big| \frac{1}{n},1,1\right) \mathcal{S}_3\left(\alpha(\sigma_0) +2+\frac{1}{n}\Big| \frac{1}{n},1,1\right)  \\
& \times \prod_{\rho \in {\rm weight}} \mathcal{S}_3^{-1}\left(m+\frac{1}{2} \left( \frac{1}{n}+2 \right)+\rho(\sigma_0) \Big|\frac{1}{n},1,1\right) \ ,
\end{aligned}
\end{align}
where $\CS_3$ is the triple sine function
\begin{align}
\CS_3 (x| \omega_1, \omega_2, \omega_3) =  \prod_{p,q,r\ge 0} \left( p\omega_1 + q\omega_2 + r\omega_3 + x\right) \left( (p+1)\omega_1 + (q+1)\omega_2 + (r+1)\omega_3 - x\right) \ .
\end{align}
Here we take the mass of hypermultiplet as $m$, and
$I_0$ is the classical contribution at the localization fixed point\footnote{Here, there is an imaginary unit in front of the Chern-Simons level $k$ because of our anti-hermitian convention for Lie algebras.}
\begin{align}
\begin{aligned}
I_0 &= -\frac{2}{g^2}\int d^5 x \sqrt{g} \, (2C - 8 t^{IJ}t_{IJ}) \text{Tr}\, \sigma_0^2 +  \frac{ik}{24\pi^2} \int \kappa \wedge d\kappa \wedge d\kappa \,\text{Tr}\,\sigma_0^3\\
	&= - \frac{8\pi^3 n}{g^2}\text{Tr}\, \sigma_0^2 +  \frac{ik\pi n}{3} \text{Tr}\,\sigma_0^3\ ,
\end{aligned}
\end{align}
where we used the volume of the $n$-covering five-sphere in the second equality.\footnote{Note that our convention leads to $\kappa\wedge d\kappa \wedge d\kappa = 8 \text{vol}(M_5)$.}
There are also instanton contributions that we will neglect in the large-$N$ limit.

\subsection{Large-$N$ limit of supersymmetric R{\'e}nyi entropy}
Consider $\CN=1$ supersymmetric $USp(2N)$ gauge theories with $N_f$ flavors and a single hypermultiplet in the antisymmetric representation.
In the large-$N$ limit, the free energy $F=-\log Z$ on the $n$-covering five-sphere is approximated by the saddle point $\sigma \to   -i N^{1/2} x$ with the density
\cite{Jafferis:2012iv,Alday:2014bta}
\begin{align}\label{density}
\rho (x) = \frac{2}{x_\ast^2} x \ ,\qquad x \in \left[ 0, x_\ast\equiv \frac{\omega_1 + \omega_2 + \omega_3}{\sqrt{2(8-N_f)}}\right] \ .
\end{align}
The free energy in the leading order of the large-$N$ is
\begin{align}\label{FreeEnergy}
F = \frac{(\omega_1 + \omega_2 + \omega_3)^3}{27 \omega_1 \omega_2 \omega_3} F_{\BS^5} = \frac{(2n+1)^3}{27n^2} F_{\BS^5}\ ,
\end{align}
where $F_{\BS^5}$ is for a round five-sphere
\begin{align}
F_{\BS^5} = - \frac{9\sqrt{2} \pi N^{5/2}}{5\sqrt{8-N_f}} \ .
\end{align}
Using the definition \eqref{SRE_def}, the supersymmetric R{\'e}nyi entropy in the large-$N$ limit is\footnote{Note that $S_1^\text{susy} = S_1$ because there is no $R$-symmetry flux in the $n\to 1$ limit.}
\begin{align}\label{SRE_largeN}
S_n^\text{susy} = - \frac{19n^2 + 7n +1}{27n^2} F_{\BS^5} = \frac{19n^2 + 7n +1}{27n^2} S_1 \ .
\end{align}

It is worth mentioning that the ratio of the supersymmetric R{\' e}nyi entropy $H_n \equiv S^\text{susy}_n/S_1$ in the large-$N$ limit satisfies the inequalities
\begin{align}
\begin{aligned}
	\partial_n H_n &\le 0 \ , \\
	\partial_n \left( \frac{n-1}{n} H_n \right) &\ge 0 \ , \\
	\partial_n ((n-1)H_n) &\ge 0 \ , \\
	\partial_n^2 ((n-1)H_n) &\le 0 \ ,
\end{aligned}
\end{align}
that are satisfied by the usual R{\'e}nyi entropy \cite{zyczkowski2003renyi} and the supersymmetric R{\'e}nyi entropy in three dimensions \cite{Nishioka:2013haa}.

\subsection{Adding Wilson loop}
The supersymmetric Wilson loop in a representation $\mathfrak R$ of the gauge group is 
\begin{align}
	W_{\mathfrak R} = \frac{1}{\text{dim} {\mathfrak R}} \text{Tr}_{\mathfrak R}\, \CP \exp \left[ -\oint ds\, 
		( A_\mu \dot x^\mu (s) - i \sigma |\dot x(s)|) 
		\right] \ .
\end{align}
This is invariant under the supersymmetry transformation if the contour of the loop is the same as the orbit of the Killing vector, namely, $\dot x^\mu(s)/|\dot x(s)| = v^\mu$.
The variation of the entanglement entropy due to the Wilson loop is considered by \cite{Lewkowycz:2013laa} and those of the supersymmetric R{\'e}nyi entropies are \cite{Nishioka:2014mwa}
\begin{align}\label{Wilson_RE}
S_{W,n}^\text{susy} = \frac{1}{n-1}\left( n \log|\langle W_{\mathfrak R} \rangle_1| - \log|\langle W_{\mathfrak R} \rangle_n|\right) \ ,
\end{align}
where $\langle \cdot \rangle_n$ stands for the expectation value taken on the $n$-covering sphere.

Let us consider a Wilson loop in the fundamental representation wrapped on $\tau$ direction. 
This configuration is BPS for arbitrary real $n$.
In the large-$N$ limit, the expectation value can be approximated at the saddle point
\begin{align}
\langle W \rangle_n = \int_0^{x_\ast} dx\, \rho(x) \, e^{2\pi n (N^{1/2}x + O(1))} \ ,
\end{align}
with the density \eqref{density}, resulting in
\begin{align}
\log \langle W\rangle_n = \sqrt{\frac{2}{8-N_f}} \,\pi (2n+1) N^{1/2} \ .
\end{align}
Then, the definition \eqref{Wilson_RE} yields the variation of the supersymmetric R{\'e}nyi entropy independent of $n$
\begin{align}\label{Wilson_SRE}
S_{W, n}^\text{susy} = \sqrt{\frac{2}{8-N_f}} \,\pi\, N^{1/2} \ .
\end{align}

\section{Gravity dual of supersymmetric R{\' e}nyi entropy}
The gravity dual of a squashed five-sphere with $SU(3)\times U(1)$ symmetry has been constructed in \cite{Alday:2014rxa,Alday:2014bta} using Romans $F(4)$ supergravity in six dimensions \cite{Romans:1985tw}.
We follow the conventions of \cite{Alday:2014rxa,Alday:2014bta} below.

Instead of finding a solution dual to the resolved five-sphere, 
we map the $n$-covering of a five-sphere to the hyperbolic space $\BS^1 \times \BH^4$ with the metric
\begin{align}
ds^2 = d\hat\tau^2 + du^2 + \sinh^2 u \,ds_{\BS^3}^2 \ ,
\end{align}
by a conformal transformation \cite{Casini:2011kv}.
Here $\hat\tau\equiv n\tau$ is the rescaled circle direction with the periodicity $\hat \tau \sim \hat \tau + 2\pi n$.
We will look for an asymptotically Euclidean AdS$_6$ solution whose boundary is $\BS^1 \times \BH^4$ in the Romans theory.

\subsection{Romans $F(4)$ supergravity}
The bosonic part of the Romans $F(4)$ supergravity includes the metric $g_{\mu\nu}$, a dilaton $\phi$, a one-form field $A$, a two-form field $B$ and $SU(2)$ gauge field $A^i$ $(i=1,2,3)$.
The Euclidean action is\footnote{We rescale the fields so that the gauge coupling $g$ appears as an overall factor.}
\begin{align}
\begin{aligned}
I = - \frac{1}{16\pi G_N g^4} \int & \Bigg[ R\ast 1- 4 X^{-2} dX \wedge \ast dX - \left( \frac{2}{9}X^{-6} - \frac{8}{3}X^{-2} - 2X^2\right) \ast 1 \\
	&\qquad - \frac{1}{2}X^{-2} \left( F\wedge \ast F + F^i \wedge \ast F^i\right) - \frac{1}{2}X^4 H \wedge \ast H \\
	&\qquad - i B\wedge \left( \frac{1}{2}dA \wedge dA + \frac{1}{3}B \wedge dA + \frac{2}{27} B\wedge B + \frac{1}{2}F^i \wedge F^i \right)\Bigg] \ ,
\end{aligned}
\end{align}
where $X\equiv e^{-\phi/\sqrt{8}}$, $F = dA + \frac{2}{3}B$, $F^i = dA^i - \frac{1}{2} \varepsilon_{ijk}A^j \wedge A^k$ and $H = dB$.
Since the gauge coupling $g$ appears only in the overall factor, we set $g=1$ in the following.
The Hodge duality is defined by
\begin{align}
\alpha \wedge \ast \beta = \frac{1}{p!}\alpha_{\mu_1\cdots \mu_p} \beta^{\mu_1\cdots \mu_p} \ast 1 \ ,
\end{align}
for $p$-forms $\alpha$ and $\beta$.

The equations of motion are
\begin{align}\label{EOM_Gauge}
\begin{aligned}
d (X^4 \ast H) &= \frac{i}{2} F\wedge F + \frac{i}{2}F^i\wedge F^i + \frac{2}{3} X^{-2} \ast F \ , \\
d (X^{-2} \ast F) &= -i F\wedge H \ , \\
D (X^{-2} \ast F^i) &= - i F^i \wedge H \ , \\
d (X^{-1} \ast dX) &= - \left( \frac{1}{6} X^{-6} - \frac{2}{3} X^{-2} + \frac{1}{2} X^2\right) \ast 1 \\
	&\qquad - \frac{1}{8} X^{-2} (F \wedge \ast F + F^i \wedge \ast F^i ) + \frac{1}{4} X^4 H \wedge \ast H \ , 
\end{aligned}
\end{align}
where the covariant derivative $D$ in the third line acts as $D \omega^i = d \omega^i - \epsilon_{ijk} A^j \wedge \omega^k$, 
and the Einstein equation is
\begin{align}\label{EOM_Einstein}
\begin{aligned}
 R_{\mu\nu} &= 4 X^{-2} \partial_\mu X \partial_\nu X + \left( \frac{1}{18} X^{-6} - \frac{2}{3} X^{-2} - \frac{1}{2} X^2\right) g_{\mu\nu} + \frac{1}{4} X^4 \left( H_{\mu\rho\sigma} H_\nu^{~\rho\sigma} - \frac{1}{6} H^2 g_{\mu\nu} \right) \\
 &\qquad + \frac{1}{2} X^{-2} \left( F_{\mu\rho}F_\nu^{~\rho} - \frac{1}{8} F^2 g_{\mu\nu}\right) + \frac{1}{2} X^{-2}\left( F^i_{\mu\rho}F_\nu^{i~\rho} - \frac{1}{8} (F^i)^2 g_{\mu\nu}\right) \ .
\end{aligned}
\end{align}

The gauge symmetry $A\to A + 2\lambda/3$, $B\to B - d\lambda$ allows us to set $A=0$. 
Also, the second equation of \eqref{EOM_Gauge} follows from the first one by taking the exterior derivative \cite{Alday:2014bta} .

\subsection{Holographic free energy}\label{ss:holographicFE}
The holographic free energy is the on-shell action on a manifold $\BM_6$ with boundary $\partial \BM_6$
\begin{align}
I_\text{tot} = I_\text{bulk} + I_\text{bdy} + I_\text{GH} + I_\text{c.t.} \ ,
\end{align}
where $I_\text{bulk}$ and $I_\text{bdy}$ are the bulk and the boundary on-shell actions
\begin{align}
\begin{aligned}
I_\text{bulk} &= \frac{1}{16\pi G_N} \int_{\BM_6}\left[ \frac{4}{9}X^{-2} (2+3X^4) \ast 1 + \frac{1}{3}X^{-2} F^i \wedge \ast F^i + \frac{i}{3} B\wedge F^i \wedge F^i \right] \ , \\
I_\text{bdy} &= \frac{1}{16\pi G_N} \int_{\partial \BM_6} \left[ \frac{2}{3}X^{-1}\ast dX + \frac{1}{3}X^4 B\wedge\ast H \right]\ .
\end{aligned}
\end{align}
$I_\text{GH}$ is the Gibbons-Hawking term added at the boundary $\partial \BM_6$ with the induced metric $h$
\begin{align}
I_\text{GH} = - \frac{1}{8\pi G_N} \int_{\partial \BM_6}  K \ast_h 1  \ ,
\end{align}
where $\ast_h$ is the Hodge dual with respect to $h$ and $K\equiv \nabla_\mu n^\mu$ is the trace of the extrinsic curvature  with the unit normal vector $n^\mu$ to $\partial \BM_6$.
The last one is the counterterm needed to make the action finite  \cite{Alday:2014rxa,Alday:2014bta} 
\begin{align}
\begin{aligned}
I_\text{c.t.} &= \frac{1}{8\pi G_N} \int_{\partial \BM_6} \Bigg[- \frac{3}{4\sqrt{2}} F^i \wedge \ast_h F^i + (\text{terms including} \, B)\\
	&\quad +\left\{ \frac{4\sqrt{2}}{3} + \frac{1}{2\sqrt{2}} R(h) + \frac{3}{4\sqrt{2}}R(h)_{mn}R(h)^{mn} - \frac{15}{64\sqrt{2}}R(h)^2 + \frac{4\sqrt{2}}{3}(1-X)^2\right\}\ast_h 1 \Bigg] \ .
\end{aligned}
\end{align}

\subsection{BPS charged topological AdS black hole}

The reasonable ansatz for the metric that preserves the boundary symmetry $SO(1,4)\times U(1)$ is
\begin{align}
ds^2_6 = h^{1/2}(r)f^{-1}(r) dr^2 + h^{-3/2}(r)f(r) d\hat\tau^2 + h^{1/2}(r)r^2\left[ du^2 + \sinh^2 u \,ds_{\BS^3}^2\right] \ .
\end{align}
Since the background fields break $SU(2)_R$ symmetry to the subgroup $U(1)$ in the dual field theory, we correspondingly assume $A^1=A^2=B=0$ and 
\begin{align}
\begin{aligned}
X &= X(r)  \ , \qquad A^3 = a(r) d\hat\tau \ .
\end{aligned}
\end{align}

In \cite{Cvetic:1999un}, AdS black hole solutions were constructed with the fields given by
\begin{align}\label{AdSBH}
\begin{aligned}
X(r) &= h^{-1/4}(r) \ , \qquad & a(r) &= \sqrt{2} (1-h^{-1}(r)) \coth\beta \ , \\
f(r) &= \frac{2}{9}r^2 h^2(r) - \frac{\mu}{r^3} - 1\ , \qquad & h(r) &= 1 - \frac{\mu \sinh^2 \beta}{r^3} \ .
\end{aligned}
\end{align}
This solution becomes the supersymmetric AdS$_6$ topological black hole when $\mu=0$, but is no longer BPS for $\mu >0$.
Actually, the BPS solution takes very close form to the above solution
\begin{align}\label{BPS_BH}
\begin{aligned}
X(r) &= h^{-1/4}(r) \ , \qquad & a(r) &= \sqrt{2} (1-h^{-1}(r))  \ , \\
f(r) &= \frac{2}{9}r^2 h^2(r) - 1\ , \qquad & h(r) &= 1 - \frac{\sqrt{2} \sinh^2 (3\gamma/2)}{r^3} \ .
\end{aligned}
\end{align}
One can check that the above solution satisfy the integrability equation of Killing spinor equation of the Euclidean Romans theory given in (A.5) of \cite{Alday:2014bta}.
This BPS black hole is asymptotically AdS space with the AdS radius $L_\text{AdS} = \frac{3}{\sqrt{2}g}$ and has a horizon at the largest root $r=r_H(\gamma)$ of $f(r_H(\gamma))=0$ depending on the parameter $\gamma$
\begin{align}
r_H(\gamma) = \frac{2\cosh \gamma + 1}{3} L_\text{AdS} \ .
\end{align}
The temperature and the entropy of the black holes are 
\begin{align}
T = \frac{2\cosh \gamma -1}{2\pi L_\text{AdS}} \ .
\end{align}
and
\begin{align}
S_\text{BH} = \frac{ L_\text{AdS}\, r_H^3(\gamma)}{4G_N}\, \text{vol}(\BH^4) \ ,
\end{align}
where $\text{vol}(\BH^4)$ is the reguralized volume of the four-dimensional unit hyperbolic space
\begin{align}
\text{vol}(\BH^4) = \frac{4\pi^2}{3} \ .
\end{align}
Since the period of the circle in the boundary has $2\pi n$ periodicity, we take the parameter $\gamma$ to be 
\begin{align}
\cosh \gamma = \frac{n+1}{2n} \ .
\end{align}
Using the renormalized action in section \ref{ss:holographicFE}, the holographic free energy becomes
\begin{align}
I_\text{tot} = - \frac{\pi^2}{4G_N} \frac{(2n+1)^3}{n^2} \ .
\end{align}
Comparing this result with the free energy on the $n$-covering five-sphere \eqref{FreeEnergy}, they completely agree due to the relation $G_N = - 15\pi \sqrt{8-N_f}/ (4\sqrt{2}N^{5/2})$ \cite{Jafferis:2012iv,Alday:2014bta}.
Therefore, the holographic supersymmetric R{\' e}nyi entropy also agrees with the large-$N$ result \eqref{SRE_largeN}. 


\subsection{Holographic Wilson loop}
Consider a Wilson loop in the fundamental representation wrapped along $\tau$ direction at the origin of $\BH^4$. 
It is dual to the fundamental string in the AdS$_6$ space \cite{Maldacena:1998im,Rey:1998ik} extending from the boundary $r=\infty$ and terminating at the horizon $r=r_H$.
The expectation value is given by
\begin{align}\label{W=S}
	\log \langle W\rangle = -S_\text{string} \ ,
\end{align}
where $S_\text{string}$ is the string world sheet action in a target space with the metric $ds^2_\text{string} = G_{\mu\nu} dx^\mu dx^\nu$
\begin{align}
S_\text{string} = \frac{1}{2\pi \alpha'} \int d^2 \xi \sqrt{\det \, G_{\mu\nu} \partial_\alpha x^\mu \partial_\beta x^\nu} \ .
\end{align}
Uplifting the solution to the massive IIA supergravity, the scalar field $X$ becomes the dilation field which distinguishes the string frame from the Einstein frame as $ds^2_\text{string} = X^{-2} ds_\text{Einstein}^2$.
Taking into account this effect and choosing the world sheet coordinates to be $\xi^1 = \tau, \xi^2 = r$, the string action becomes
\begin{align}
S_\text{string} = \frac{n L_\text{AdS}}{\alpha '} (r_\infty - r_H) = - \frac{2n +1}{3 \alpha '} L_\text{AdS}^2 \ .
\end{align}
From the relation \eqref{W=S} and the definition \eqref{Wilson_RE}, we obtain the variation of the supersymmetric R{\'e}nyi entropy 
\begin{align}\label{Wilson_HSRE}
S_{W,n}^\text{susy} = \frac{L^2_\text{AdS}}{3\alpha'} \ ,
\end{align}
which is independent of the parameter $n$.
If the solution is dual to $\CN=1$ $USp(2N)$ superconformal theories, the AdS radius is fixed to be $L_\text{AdS}^2/\alpha ' = 3\sqrt{2} \pi N^{1/2}/\sqrt{8-N_f}$ \cite{Bergman:2012kr}
and \eqref{Wilson_HSRE} agrees with the field theory result \eqref{Wilson_SRE}.

 \vspace{1.3cm}
 \centerline{\bf Acknowledgements}
We are grateful to T.\,Nosaka, S.\,Pufu and M.\,Yamazaki for valuable discussions.
The work of N.\,H. was supported in part by JSPS Research Fellowship for Young Scientists. 
The work of T.\,U. was supported in part by the National Science Foundation under Grant No.\,NSF\, PHY-25915. 

\appendix

\section{Convention}\label{ss:convention}

\begin{itemize}
\item
The Greek indices $\mu,\nu,\cdots$ are for the space-time indices.
\item
The Roman indices $a,b,\cdots$ are for the local Lorentz frame.
To translate them into the space-time indices, we use vielbein $e^\mu_a$. For instance, $v^{\mu \nu} = v^{ab} e^\mu_a e^\nu_b$.
\item
The capital Roman indices $I,J,\cdots$ stand for the $SU(2)_R$ symmetry, $I,J =1,2$.
\item
The capital Roman indices $A,B,\cdots$ stand for the $Sp(2r)$ flavor symmetry, $A,B=1,2,\cdots, 2r$.
\item
The spinor indices are denoted by the Greek indices $\alpha,\beta,\cdots$.
\item
The generators of the Lie algebra $T^A$ of a Lie group is anti-hermitian: $T_A^\dagger = -T_A$ satisfying $[T_A, T_B] = - f_{AB}^{~~C} T_C$ with the structure constant $f_{AB}^{~~C}$ and the Killing form 
$\text{Tr} (T_A T_B) = -\frac{1}{2}\delta_{AB}$.
\item All spinors are Grassmann odd except Killing spinors that are treated as Grassmann \emph{even} variables.
\item The gauge transformation $\delta_G$ for a gauge field $A$ defined by
\begin{align}
\delta_G (\epsilon) A = d\epsilon - [A, \epsilon] \ ,
\end{align}
and for a matter field $\Phi$ by
\begin{align}
\delta_G (\epsilon) \Phi &= \epsilon \Phi \ .
\end{align}
\end{itemize}

\subsection{Gamma matrices and spinors}
In the Euclidean five dimensions, the gamma matrices satisfying the Clifford algebra
\begin{align}
\{ \Gamma_a, \Gamma_b \} = 2\delta_{ab} \ , \qquad (a,b=1,2,\cdots,5) \ ,
\end{align}
are $4\times 4$ matrices.
We use the gamma matrices with multiple indices 
\begin{align}
\Gamma^{a_1a_2\cdots a_n} \equiv \frac{1}{n!} \sum_{\sigma\in S_n}(-1)^{|\sigma|}\Gamma^{a_{\sigma(1)}}\Gamma^{a_{\sigma(2)}}\cdots \Gamma^{a_{\sigma(n)}} \ ,
\end{align}
where $|\sigma|$ is the order of an element $\sigma$ of the permutation group $S_n$.
With this notation, the product of the gamma matrices obeys the relation
\begin{align}
\Gamma_{1\cdots 5} = -{\bf 1}_4 \ .
\end{align}
The charge conjugation matrix $\CC$ satisfies
\begin{align}
\CC^T = -\CC \ , \qquad \CC^\dagger \CC = 1 \ , \qquad \Gamma_a^T = \CC \Gamma_a \CC^{-1} \ .
\end{align}

We introduce a spinor $\psi^\alpha$ with upper index $\alpha$. Then the positions of indices of the matrices are
\begin{align}
(\Gamma_a)^\alpha_{~\beta} \ , \qquad  \CC_{\alpha\beta} \ ,\qquad \CC^{\alpha\beta}\CC_{\beta\gamma} = -\delta_\gamma^\alpha \ , \qquad (\CC^{\alpha\beta})^\ast = \CC^{\alpha\beta} = \CC_{\alpha\beta} \ .
\end{align}
We raise and lower the spinor index by the charge conjugation matrix
\begin{align}
\psi_\alpha = \psi^\beta \CC_{\beta\alpha} \ , \qquad \psi^\alpha = \CC^{\alpha\beta} \psi_\beta \ ,
\end{align}
and the contraction is always defined by
\begin{align}
\chi\psi \equiv \chi_\alpha \psi^\alpha \ .
\end{align}
Spinors have an $SU(2)_R$ index $I$ $(I=1,2)$ which can be raised or lowered by the skew-symmetric matrix $\epsilon_{IJ}$:
\begin{align}
\psi^I = \epsilon^{IJ}\psi_J \ , \qquad \psi_I = \psi^J \epsilon_{JI} \  , \qquad 
\epsilon_{IJ}=\epsilon^{IJ} = (i\sigma_2)_{IJ} \ .
\end{align}

We can impose the $SU(2)$-Majorana condition on spinors with the charge conjugation matrix
\begin{align}
 (\psi_{I}^\alpha)^\ast  = \epsilon^{IJ} \psi_J^{\beta}\, \CC_{\beta\alpha} \ .
\end{align}
If we denote spinors as $\epsilon= (\epsilon^1, \epsilon^2)^T$, the condition yields
\begin{align}
\epsilon^2 = \CC^\ast (\epsilon^1)^\ast \ .
\end{align}

The bilinear terms of spinors with $SU(2)_R$ indices 
behave under the exchange of the two spinors as
\begin{align}
\psi^I \Gamma^{a_1a_2\cdots a_n} \chi^J = t_n\,  \chi^J \Gamma^{a_1a_2\cdots a_n} \psi^I \ , \qquad 
t_n = \left\{\begin{array}{ll}
+1\ , & n=0,1,4,5 \\
-1\ , & n = 2,3
\end{array}\right. \ .
\end{align}
When the indices of $SU(2)_R$ are contracted, there appears a minus sign in the right hand side.

For triple spinors $(\xi,\eta,\psi)$, the Fierz identities hold
\begin{align}
\begin{aligned}
\xi^\alpha (\eta \psi) &= \frac{(-1)^s}{4} \left[ \psi^\alpha (\eta\xi) + (\Gamma^\mu\psi)^\alpha (\eta \Gamma_\mu \xi) - \frac{1}{2}(\Gamma^{\mu\nu}\psi)^\alpha (\eta\Gamma_{\mu\nu}\xi) \right] \ , \\
(\Gamma^\mu \xi)^\alpha (\eta\Gamma_\mu\psi)&= -2 \psi^\alpha (\xi\eta) - 2\eta^\alpha (\psi \xi) - \xi^\alpha (\eta\psi) \ , \\
(\Gamma^{\mu\nu}\xi)^\alpha (\eta \Gamma_{\mu\nu}\psi) &= 4\psi^\alpha (\xi\eta) - 4\eta^\alpha (\psi\xi) \ , 
\end{aligned}
\end{align}
where $s=1$ for Grassmann odd spinors and $s=0$ for Grassmann even spinors.

\subsection{Spin connection and Lie derivatives}
The spin connection for a given vielbein is defined by
\begin{align}
\omega_\mu^{~ab} = e_\nu^a \nabla_\mu e^{\nu b} \ ,
\end{align}
and the Riemann tensor is given using the spin connection by
\begin{align}
R_{\mu\nu}^{~~ab} = \partial_\mu \omega_\nu^{~ab} - \partial_\nu \omega_\mu^{~ab} + \omega_{\mu}^{~ac}\omega_{\nu c}^{~b} - \omega_{\nu}^{~ac}\omega_{\mu c}^{~b} \ .
\end{align}
The Ricci scalar is then 
\begin{align}
R = e_{a}^\nu e_b^\mu R_{\mu\nu}^{~~ab} \ .
\end{align}
This convention yields a negative Ricci curvature for a round sphere.

The covariant derivatives for spinors are defined by
\begin{align}
\nabla_\mu \zeta = \partial_\mu\zeta + \frac{1}{4}\omega_{\mu}^{~ab}\Gamma_{ab}\zeta \ .
\end{align}

\section{Omega deformation parameters}\label{app:OmegaBG}
In this appendix, we justify the identification of the resolved space with the squashed five-sphere with the squashing parameters  \eqref{triplesines}. 
We relabel the phases of $z_i$ by $\phi_i$
\begin{align}
\phi_1&=\tau \ , \quad \phi_2=\frac{1}{2}(\chi+\xi)\ , \quad \phi_3=\frac{1}{2}(\chi-\xi) \ ,
\end{align}
then 
\begin{align}
\begin{aligned}
\partial_{\phi_1} &= \partial_\tau \ , \\
\partial_{\phi_2} &= \partial_\chi+\partial_\xi \ , \\
\partial_{\phi_3} &= \partial_\chi - \partial_\xi \ .
\end{aligned}
\end{align}
Since the Hopf fibration rotates all $z_i$'s by a same angle, the direction of this fibration can be arranged into $w^\mu \partial_\mu = a \sum_i \partial_{\phi_i}$, and remaining part $\partial_{\tau''} = a_i \partial_{\phi_i}$ of Killing vector $v= \frac{1}{n}\partial_\tau+2\partial_\chi=w^\mu \partial_\mu+\partial_{\tau''}$ translates the base manifold of the Hopf fibration. 
Here we restrict the parameters as $a_1+a_2+a_3=0$ so that $\partial_{\tau''}$ acts only base manifold.
These parameters are fixed to be 
\begin{align}
\begin{aligned}
a_1&=\frac{2}{3n}(1-n)\ , \qquad & a_2&=a_3= -\frac{1}{3n}(1-n) \ ,\\
\partial_{\tau''} &= \frac{2}{3n}(1-n)(\partial_\tau-\partial_\chi)\ , \qquad &w^\mu \partial_\mu &= \frac{2n+1}{3n}(\partial_\tau+2\partial_\chi)\ ,
\end{aligned}
\end{align}
and their directions are parametrized by $\partial_{\tau'}=w^\mu\partial_\mu$ and $\tau''$ by 
\begin{align}
\begin{aligned}
\tau'&=\frac{n}{2n+1}(\tau+\chi)\ ,\qquad &\tau'' &= \frac{n}{2(1-n)}(2\tau-\chi) \ , \\
\tau&=\frac{1}{3n}((1+2n)\tau'+2(1-n)\tau'')\ ,\qquad &\chi&=\frac{2}{3n}((1+2n)\tau'-(1-n)\tau'')\ .
\end{aligned}
\end{align}
The metric of the base four-dimensional manifold $ds_{\mathbb{M}_4}^2$ are derived by square completion in $d\tau'$, the direction of the Hopf fibration. 
In fact, by using 
\begin{align}
w_\mu dx^\mu = \left( \frac{1+2n}{3n} \right)^2 (n^2 \sin^2 \theta +\cos ^2 \theta) \left[ d\tau' + \frac{2(2n^2 \sin^2 \theta -\cos ^2 \theta)d\tau''+3n\cos^2\theta\cos\phi d\xi}{2(1+2n)(n^2 \sin^2 \theta +\cos ^2 \theta)} \right]\ ,
\end{align}
the metric of the original manifold turns out to be
\begin{align}
\begin{aligned}
ds^2&= \frac{d\theta^2}{f_n(\theta)}+n^2 \sin^2 \theta d\tau^2 +\frac{1}{4} \cos^2 \theta (d\phi^2+d\xi^2+d\chi^2+2\cos \phi d\xi d\chi ) \ , \\
&= \left( \frac{3n}{1+2n} \right)^2 \frac{(w_\mu dx^\mu)^2}{n^2 \sin^2 \theta +\cos ^2 \theta}\ , \\
& + \frac{d\theta^2}{f_n} +\frac{1}{4}\cos^2 \theta (d\phi^2 +\sin^2\phi d\xi^2) + \frac{\sin^2\theta\cos^2\theta}{n^2 \sin^2 \theta +\cos ^2 \theta}\left[ (1-n)d\tau'' - \frac{n}{2}\cos \phi d\xi \right]^2\ .
\end{aligned}
\end{align}
$ds_{\mathbb{M}_4}^2$ is the third line of this metric. The squashing parameters in \eqref{triplesines} are read from the one-loop determinant in instanton configurations \eqref{saddlepts}. 
For a while, we take the translation along $\xi$ direction with charge $\delta$ into the translation of base $\mathbb{M}_4$ generated by the Killing vector. 
This is expressed as $a_2-a_3=\delta \neq 0$, and finally $\delta$ is taken to be zero\footnote{This manipulation is expected to be justified by taking another sixth direction $\BS^1$ and twisting this $U(1)$ with this $\BS^1$, for instance.}. 
Then the Killing vector which acts on $\mathbb{M}_4$ is $\delta\partial_\xi + \partial_{\tau''}$. 
The fixed points of this translation are $(z_1,z_2,z_3)=(n,0,0)\ ,(0,1,0)\ ,(0,0,1)$. 
In following, we determine the squashing parameters in \eqref{triplesines} from the configurations around these fixed points.

Firstly we consider the metric around $x_1 \equiv (n,0,0)$. 
The distance from this point is parametrized by $\epsilon  \ll 1$ where $\theta=\frac{\pi}{2}+\epsilon$. 
Hence in the first order in $\epsilon$, the metric near $(n,0,0)$
\begin{align}
ds^2 \simeq \left( \frac{3}{1+2n} \right)^2 (w_\mu dx^\mu)^2 + d\epsilon^2 + \frac{\epsilon^2}{4} (d\phi^2 + \sin^2 \phi d\xi^2)+\frac{1}{n^2}\left[ (1-n)d\tau''-\frac{n}{2}\cos\phi d\xi \right]^2 \ ,
\end{align}
is interpreted as $\BS^1$-fibration over a two-dimensional complex plane $(\xi^1_1, \xi^1_2)$, 
 $ds^2= \left( \frac{3}{1+2n} \right)^2 (w_\mu dx^\mu)^2+\sum_{i=1}^2 |d\xi^1_i|^2$, with
\begin{align}
\begin{aligned}
\xi_1^1 &= \frac{nz_2}{z_1} \simeq -\epsilon \cos \frac{\phi}{2} e^{i(\frac{\xi}{2}-\frac{1-n}{n}\tau'')} \ , \\
\xi_2^1 &= \frac{nz_3}{z_1} \simeq -\epsilon \sin \frac{\phi}{2} e^{-i(\frac{\xi}{2}+\frac{1-n}{n}\tau'')}\ .
\end{aligned}
\end{align}
This two-dimensional complex plane $\mathbb{M}_4$ is Omega-deformed by the Killing vector $\delta\partial_\xi + \partial_{\tau''}$ with Omega-deformation parameters that can be read off from the rotated angle of $\xi^1_i$ by the vector:
\begin{align}
\begin{aligned}
\epsilon_1^1 &=-\frac{1-n}{n}+\frac{\delta}{2} \ , \\
\epsilon_2^1 &=-\frac{1-n}{n}-\frac{\delta}{2}\ .
\end{aligned}
\end{align}
In this sense, the base manifold $\mathbb{M}_4$ around this fixed point is interpreted as Omega-deformed four-dimensional plane $\mathbb{R}^4_{-\frac{1-n}{n}+\frac{\delta}{2},-\frac{1-n}{n}-\frac{\delta}{2}}$.\\
In the same manner, we parametrize the configuration around $x_2 \equiv (0,1,0)$ by $\theta=\epsilon, \phi=\epsilon'$ where $\epsilon, \epsilon' \ll 1$. 
The metric of the original manifold is written in terms of
\begin{align}
\begin{aligned}
\xi_1^2&= \frac{z_1}{z_2} \simeq n\epsilon^{i(-\frac{\xi}{2}+\frac{1-n}{n}\tau'')} \ , \\
\xi_2^2&= \frac{z_3}{z_2} \simeq \frac{\epsilon'}{2} e^{-i\xi}\ ,
\end{aligned}
\end{align}
as $ds^2= \left( \frac{3n}{1+2n} \right)^2 (w_\mu dx^\mu)^2+\sum_{i=1}^2 |d\xi^2_i|^2$. 
Also this base manifold is an Omega-deformed four-dimensional plane $\mathbb{R}^4_{\frac{1-n}{n}-\frac{\delta}{2},-\delta}$.
Finally, the coordinates around the third fixed point $x_3 \equiv (0,0,1)$ are represented by $\theta=\epsilon, \phi=\pi+\epsilon'$ where $\epsilon, \epsilon' \ll 1$. 
Again, the metric $ds^2= \left( \frac{3n}{1+2n} \right)^2 (w_\mu dx^\mu)^2+\sum_{i=1}^2 |d\xi^3_i|^2$ around this point is written as a four-dimensional plane $\mathbb{R}^4_{\frac{1-n}{n}+\frac{\delta}{2},\delta}$ coordinated by
\begin{align}
\begin{aligned}
\xi_1^3&= \frac{z_1}{z_3} \simeq n\epsilon^{i(\frac{\xi}{2}+\frac{1-n}{n}\tau'')} \ , \\
\xi_2^3&= \frac{z_2}{z_3} \simeq -\frac{\epsilon'}{2} e^{i\xi} \ ,
\end{aligned}
\end{align}
whose Omega deformation parameters are obtained from their charges.

From above results, we read a one-loop determinant $Z_{{\rm one-loop}}= \prod_j w_j^{-\frac{1}{2}c_j}$ around the saddle point configuration \eqref{saddlepts} from ${\rm ind}D_{10}=\sum_j c_j e^{w_j}$. 
Since the indices of vector multiplets in five-dimensional $\mathcal{N}=1$ theory are related to those of the self-dual complex ${\rm ind}D_{{\rm SD}}$ \cite{Gomis:2011pf}, which are represented by the indices ${\rm ind}\bar{D}$ of the Dolbeault complex as
\begin{align}
{\rm ind}D_{{\rm SD}} = \frac{1}{2} (1+e^{i(\epsilon_1+\epsilon_2)})) {\rm ind}\bar{D}\ .
\end{align}
We calculate ${\rm ind}\bar{D}$ in the following. 
The indices of Dirac complex associated with the one-loop determinant of hypermultiplets are also related to ${\rm ind}\bar{D}$. 
The index of vector bundle $V$ on a manifold $\mathbb{M}_4$ is calculated by the Riemann-Roch-Hirzebruch theorem as 
\begin{align}
{\rm ind}\bar{\partial}_V = \int_{\mathbb{M}_4} {\rm Td}(T\mathbb{M}_4) {\rm Ch}(V)\ .
\end{align}
The integral of an equivariant closed form $\alpha$ localizes to the fixed points of the isometry $d$ : $\int \alpha = \sum_{p} \frac{\alpha(x_p)}{\chi(x_p)}$ where $x_p$ are fixed points of the isometry and $\chi$ is the Euler class. 
The quantities in the right-hand side are evaluated in zero-form value at $x_p$.

\paragraph{Todd class}
The Todd class for $\mathbb{C}^2$ is $\frac{\lambda_1}{1-e^{\lambda_1}}\frac{\lambda_2}{1-e^{\lambda_2}}$ where $\lambda_{1,2}$ are Chern roots. 
These are deformed as $\lambda_i \rightarrow \lambda_i+i \epsilon_i$ by Omega-deformation parameters $\epsilon_i$. 
In particular, the zero-form value of the Todd class becomes $\frac{-\epsilon^i_1\epsilon^i_2}{(1-e^{i\epsilon^i_1})(1-e^{i\epsilon^i_2})}$.

\paragraph{Euler class}
The Euler class for an oriented four-dimensional manifold $\mathbb{M}_4$ is written in terms of their curvature: $\chi(\mathbb{M}_4)=\frac{1}{32\pi^2}\epsilon^{ijkl}R_{ij}R_{kl}$. 
In the case of $\mathbb{R}^4_{\epsilon^i_1,\epsilon^i_2}$, this is equal to $-\epsilon^i_1 \epsilon^i_2$.

\paragraph{K{\"a}hler-Hodge manifold}
When K{\"a}hler transformation defines a line bundle, its first Chern class is equivalent to its K{\"a}hler class: ${\rm Ch}(V) = e^{c_1(V)}=e^{tJ}$ where $J$ is the K{\"a}hler two-form. 
To implement the equivariance for our case, considering that the K{\"a}hler two-form is closed including the twist by $\partial_{\tau''}$, we have to add zero-form value to $J$. 
In fact, 
\begin{align}
-iJ_\epsilon = \frac{1}{2} \left( \frac{1}{n^2}dz_1 \wedge d\bar{z}_1+dz_2 \wedge d\bar{z}_2+dz_3 \wedge d\bar{z}_3 \right) + \frac{a_1}{n^2} |z_1|^2+a_2|z_2|^2+a_3|z_3|^2 \ ,
\end{align}
is closed for $d+i \partial_{\tau''}$. 
Because of this, a contribution from the $U(1)$ fibration is evaluated as ${\rm Ch}(V) = e^{c_1(V)}=e^{tJ_\epsilon}$.
At the fixed points $x_i$ $(i=1,2,3)$, the zero-form values are $a_i$.

\paragraph{One-loop determinant}
The Chern class of the base $ \mathbb{M}_4$ is derived from the zero-form values of the Chern class of $\mathbb{R}^4_{\epsilon^i_1,\epsilon^i_2}$ which is identified with the local geometry around fixed points $x_i$ of $\mathbb{M}_4$. 
We assume that contributions from nonzero instanton sectors do not dominate in the large-$N$ limit, and we are only interested in the perturbative part in the zero instanton sector. 
Under these assumptions, we can evaluate the contribution from the universal bundle as ${\rm Ch}(\mathcal{E}) = {\rm Ch}_{{\rm adj}} (e^{i \lambda})$, and we can use the indices on $\mathbb{R}^4 \times \BS^1$. 
The circumference of this $\BS^1$, which is a period of $\tau'=\frac{n}{2n+1}(\tau+\chi) = 2 \pi \frac{3n}{2n+1}$, is used for the normalization of the charge $t$ of this $\BS^1$ rotation.

Putting all together, the index for vector multiplets is
\begin{align}
\begin{aligned}
I_{{\rm vec}}= \frac{1}{2} \sum_{t=-\infty}^{\infty} &\left[  \frac{(1+e^{-\frac{2i}{n}(1-n)}) e^{it(\frac{1+2n}{3n}+\frac{2}{3n}(1-n))}  }{(1-e^{-i\frac{1-n}{n}})^2} \right. \\
 &\left. +\frac{(1+e^{\frac{i}{n}(1-n)})e^{it(\frac{1+2n}{3n}-\frac{1-n}{3n})}}{(1-e^{i\frac{1-n}{n}})(1-e^{-i\delta})} + \frac{(1+e^{\frac{i}{n}(1-n)})e^{it(\frac{1+2n}{3n}-\frac{1-n}{3n})}}{(1-e^{i\frac{1-n}{n}})(1-e^{i\delta})} \right] {\rm Ch}_{{\rm adj}} (e^{i \lambda}) \ .
\end{aligned}
\end{align}
Each term has expansions in positive and negative power series \cite{Pestun:2007rz}, and the one-loop determinant are derived from their sum.
\begin{align}
\begin{aligned}
{\rm det}_V &= \prod_{\alpha \in {\rm root}} \prod_{p,q,r \geq 0} \frac{1}{i\alpha(\lambda)} \left( \frac{p}{n}+q+r +i \alpha(\lambda) \right) \left( \frac{p+1}{n} +q+r+2+i \alpha(\lambda) \right) \ , \\
&=\prod_{\alpha \in {\rm positive \ root}} \frac{1}{(i\alpha(\lambda))^2}  \mathcal{S}_3\left(i\alpha \bigg| \frac{1}{n},1,1\right) \mathcal{S}_3\left(i\alpha +2+\frac{1}{n}\bigg| \frac{1}{n},1,1\right) \ ,
\end{aligned}
\end{align}
where we took $\delta$ to be zero. 
The triple sine function $\mathcal{S}_3(x|\omega_1,\omega_2,\omega_3)$ is defined as a
regularized infinite product $\prod_{p,q,r \geq 0} (p\omega_1+q\omega_2+r\omega_3+x)((p+1)\omega_1+(q+1)\omega_2+(r+1)\omega_3-x)$. 
Similarly, the contribution from a hypermultiplet in the representation $\mathfrak{R}$ and mass $m$ is
\begin{align}
\begin{aligned}
{\rm det}_H = \prod_{\rho \in {\rm weight}} \prod_{p,q,r \geq 0} \prod_\pm & \left[ \frac{p}{n}+q+r+m+\frac{1}{2} \left( \frac{1}{n}+2 \right)\pm i \rho(\lambda) \right]^{-\frac{1}{2}} \\
& \times\left[ \frac{p+1}{n}+q+r+2-m-\frac{1}{2} \left( \frac{1}{n}+2 \right)\pm i \rho(\lambda) \right]^{-\frac{1}{2}} \ .
\end{aligned}
\end{align}
In case that $\mathfrak{R}$ is a real representation, it simplifies to
\begin{align}
\begin{aligned}
{\rm det}_H &= \prod_{\rho \in {\rm weight}} \prod_{p,q,r \geq 0} \left[ \frac{p}{n}+q+r+m+\frac{1}{2} \left( \frac{1}{n}+2 \right) + i \rho(\lambda) \right]^{-1} \\
& \quad\quad\quad\quad \times\left[ \frac{p+1}{n}+q+r+2-m-\frac{1}{2} \left( \frac{1}{n}+2 \right) - i\rho(\lambda) \right]^{-1} \ ,\\
&= \prod_{\rho \in {\rm weight}} \mathcal{S}_3^{-1}\left(m+\frac{1}{2} \left( \frac{1}{n}+2 \right)+i\rho \bigg|\frac{1}{n},1,1\right)\ .
\end{aligned}
\end{align}
Comparing these with the one-loop partition function on the squashed five-sphere \cite{Lockhart:2012vp,Imamura:2012bm}, we 
confirm that the squashing parameters are given by \eqref{triplesines} for the resolved space.


\bibliographystyle{JHEP}
\bibliography{SRE5d}

\end{document}